\documentclass[12pt,preprint]{aastex} %single column single-spaced document
\shorttitle{Photospheric Net Currents in Vector Magnetograms}
\shortauthors{Venkatakrishnan and Tiwari}

\begin{document}

\title{On the Absence of Photospheric Net Currents in Vector Magnetograms
of Sunspots Obtained From {\it Hinode} (SOT/SP)}

\author{P.~VENKATAKRISHNAN and
SANJIV KUMAR TIWARI}
\affil{Udaipur Solar Observatory, Physical Research Laboratory,
 Dewali, Bari Road,\\ Udaipur-313 001, India}
\email{pvk@prl.res.in}
\email{stiwari@prl.res.in}

\begin{abstract}

Various theoretical and observational results have been reported regarding
the presence/absence of net electric currents in the sunspots.
The limited spatial resolution of the earlier observations perhaps
obscured the conclusions. We have
analyzed 12 sunspots observed from Hinode (SOT/SP) to clarify the issue.
The azimuthal and radial components of magnetic fields
and currents have been derived.
The azimuthal component of the magnetic field of sunspots is found to vary in
sign with azimuth. The radial component of the field also varies in magnitude
with azimuth. While the latter pattern is a confirmation of the interlocking
combed structure of penumbral filaments, the former pattern shows that the
penumbra is made up of a ``curly interlocking combed" magnetic field.
The azimuthally averaged azimuthal component is seen to decline much faster
than 1/$\varpi$ in the penumbra, after an initial increase in the umbra,
for all the spots studied. This confirms the confinement of magnetic fields
and absence of a net current for sunspots as postulated by \cite{parker96}.
The existence of a global twist for a sunspot even in the absence of
a net current is consistent with a fibril-bundle structure of the sunspot
magnetic fields.

\end{abstract}


\section{Introduction}

Sunspots have shown evidence for twist even from the time of
\cite{hale25,hale27} who postulated the hemispheric rule for the
chirality of chromospheric whirls.
This was later confirmed with a larger data set by \cite{rich41}.
Evidence for photospheric chirality could be seen in early continuum
images of sunspots, obtained with exceptional image quality.
Later, photospheric vector magnetograms showed global twist
inferred from the non-vanishing averages of the force-free parameter
(\cite{pcm94,hagi04,nand06} and references therein).
The non-force-free nature of photospheric magnetic field
in the sunspots, prompted \cite{tiw09a} to propose the signed shear
angle (SSA) as a more robust measure of the global twist of the
sunspot magnetic field.

Although, the sign of SSA matches well with the sign of the global
alpha parameter, the magnitudes are not so well correlated.
The physical significance of a globally averaged $\alpha$ parameter
rests heavily on the existence of a net current in the
photospheric sunspot magnetic field.
One way of arriving at a global $\alpha$ is by taking the
ratio of total vertical current to the total flux (integral method).
This value was found to agree with the values obtained by other
methods \citep{hagi04}.

For a monolithic sunspot
magnetic field, the global twist and net current is
expected to be well correlated by Ampere's Law.
However, the existence of a
net current is ruled out theoretically for fibril
bundles as well as for monolithic fields with azimuthal field
decreasing faster than 1/$\varpi$, where $\varpi$ is the radial
distance from the spot center \citep{parker96}.
Several attempts to resolve this problem using vector
magnetograms have not been
very conclusive so far \citep{wilk92,leka96,wheat2000}.

A resolution of this problem can be used to disentangle the relation between
global twist and the global $\alpha$ parameter. Also, the resolution is needed
to evaluate the so called hemispheric helicity rule seen in the global
$\alpha$ parameter calculated from photospheric vector magnetograms
\citep{pcm94,pcm95,hagi04,nand06}. The availability of high resolution vector
magnetograms from Hinode (SOT/SP), gives us the best opportunity so far to
address this problem. The effect of polarimetric noise is expected to be
negligible in the estimation of magnetic parameters \citep{tiw09b} from
these data.

In this Letter we obtain an expression for the net
current using a generalization of the expression obtained by \cite{parker96}.
We then proceed to measure this current from several vector magnetograms of
nearly circular sunspots. We finally discuss the results and present
our conclusions.

\section{Expression For Net Current}

Following \cite{parker96}, we consider a long straight flux bundle surrounded
by a region of field free plasma. We use the words ``field free" in the
empirical sense that there is no large scale coherent and unipolar magnetic
field surrounding the flux bundle. Also, we include the case where the bundle
can be replaced by a monolithic field. \cite{parker96} assumed azimuthal
symmetry as well as zero radial component $B_r$, of the magnetic field.
For realistic sunspot fields, we need to relax both these assumptions.

The vertical component of the electric current density consists of two terms,
viz. $-\frac{1}{\mu_0 r}\frac{\partial B_r}{\partial\psi}$ ~and~
$\frac{1}{\mu_0 r}\frac{\partial (r B_{\psi})}{\partial r}$.
We will call the first term as the ``pleat current density", $j_p$ and
the second term as the ``twist current density", $j_t$.
The net current $I_z$ within a distance $\varpi$ from the
center is then given by
\begin{equation}\label{}
 I_z(\varpi) = \int_0^{2\pi} d\psi \int_0^\varpi rdr(j_p + j_t)
\end{equation}
The $\psi$ integral over $j_p$ vanishes, while the second term yields
\begin{equation}\label{}
I_z(\varpi) = \frac{\varpi}{\mu_0} \int d\psi B_\psi(\varpi,\psi)
\end{equation}\label{}
which gives the net current within a circular region of radius
$\varpi$.

\section{The Data Sets and Analysis}

We have analyzed the vector magnetograms obtained from Solar
Optical Telescope/Spectro-polarimeter
(SOT/SP: \cite {tsun08,shim08,suem08,ichi08})
onboard Hinode \citep{kosu07}.
The calibration of data sets have been performed using the
standard ``SP\_PREP'' routine developed by B. Lites and available
in the Solar-Soft package. The prepared polarization spectra have
been inverted to obtain vector magnetic field components using an
Unno-Rachkowsky \citep{unno56,rach67}
inversion under the assumption of Milne-Eddington (ME)
atmosphere \citep{lando82,skum87}. We have used the inversion
code ``STOKESFIT" which has been kindly made available
by  T. R. Metcalf as a part of the
Solar-Soft package. We have used the newest version of this code
which returns true field strengths along with the filling factor.
The azimuth determination has inherent 180$^o$ ambiguity due
to insensitivity of Zeeman effect to orientation of
the transverse fields. Numerous techniques have been developed
and applied to resolve this problem, but not even one
guarantees a complete resolution.
The 180$^o$ azimuthal ambiguity in our data sets are
removed by using acute angle method \citep{harv69,saku85,cupe92}.

In order to minimize the noise, pixels with
transverse $(B_t)$ and longitudinal magnetic
field $(B_z)$ greater than a certain level are only analyzed.
A quiet Sun region is selected for each sunspot and
1$\sigma$ standard deviation in the three vector field
components $B_x$, $B_y$ and $B_z$ are evaluated
separately. The resultant standard deviations of $B_x$ and $B_y$
is then taken as the 1$\sigma$ noise level
for transverse field components. Only
those pixels where longitudinal and transverse
fields are simultaneously greater than twice the
above mentioned noise levels are analyzed.
The data sets with their observation details are given
in Table 1. We have treated each polarity as an individual
sunspot whenever both the polarities are observed and compact
enough to be studied.
We have studied only those spots where the polarity inversion
lines are well separated from the edge of the sunspot.

The results of the inversions yield the 3 magnetic parameters,
viz. the field strength B, the inclination to the line of sight
$\gamma$, and the azimuth $\phi$. These parameters are used
to obtain the 3 components of magnetic field in Cartesian geometry as
\begin{eqnarray}\label{}
B_z = B \cos \gamma\\
B_y = B \sin \gamma \sin \phi\\
B_x = B \sin \gamma \cos \phi
\end{eqnarray}\label{}
This vector field is transformed to heliographic coordinates
\citep{venk89} for the spots observed at viewing angle more
than $10^{\circ}$.
The transverse vector is then expressed in cylindrical geometry as
\begin{eqnarray}\label{}
B_r = \frac{1}{r}(xB_x + yB_y)\\
B_\psi = \frac{1}{r}(-yB_x + xB_y)
\end{eqnarray}\label{}
The azimuthal field $B_\psi$ is then used in equation (2) for obtaining
the value for the total vertical current within a radius $\varpi$.

We have computed ``twist angle'' for all the sunspots
using $B_\psi$ and $B_r$ as shown in Table 1.
The error in ``twist'' measurement is simply the error in
azimuth measurement.
Using the weak field approximation, we can find the azimuth $\psi$ from
$\tan 2\psi = U/Q$. From this we can estimate the error in $\psi$ as
equal to the percentage error in linear polarization measurements.
Thus, a 1\% error in polarimetry means that the error in $\psi$ equals
0.01 radians or 0.57 degrees. We have performed Monte Carlo simulations
of the effect of noise on the inversions which we plan to present in
a more detailed paper. We have verified that the error in $\psi$
is consistent with the value estimated from the weak field
approximation.

We can see in the Table 1 that the twist angles
for regular sunspots match
well with the global SSA as expected, whereas they do not match for
irregular sunspots.

\section{\textbf{The Results}}

Figure 1 shows an example of the maps of twist current,
pleat current, $B_{\psi}$ and $B_r$ for a sunspot NOAA AR 10933
which is nearly circular.
Figure 2(a) shows plots of $B_{\psi}$ and $B_r$
along with the different concentric circles around spot center.
In Figure 2(b), the spatial variation of both $B_{\psi}$
and $B_r$ are clearly seen. This variation is corresponding to
a typical circle selected in the penumbra.
The $B_r$ variation in the penumbra is a manifestation of the
interlocking combed structure \citep{ichi07,tiw09a}.
The $B_{\psi}$ variation in the penumbra shows that not only is
there an interlocking combed structure, but these structures
are curled as well. In other words, we may describe the
penumbral field as possessing a
``curly interlocking combed'' structure.
This feature of the deviation of the vector field azimuths from a
radial direction was also seen by \cite{shibu03} in the magnetic
field of a sunspot belonging to NOAA AR 8706,
using the infra-red FeI line pair at 1.56 micron.

The azimuthal averages $<B_{\psi}>$ and $<B_r>$ were obtained
at different values of $\varpi$.
Figure 2(c) shows the plots of $<B_{\psi}>$ and $<B_r>$
as a function of $\varpi$. The circles corresponding to the
selected radii are shown in the upper panel of the same figure.
The azimuth-averaged $<B_r>$ drops rapidly to a very
low value at the edge of the sunspot. This is a clear evidence
for the existence of a canopy where the field lines lift up
above the line forming region.  Figure 3 shows the plot of
log $<B_\psi>$ as a function of log $\varpi$. The slope $\delta$ of the
declining portion of this plot is 9.584,
which shows that field varies faster than 1/$\varpi$.
This can be construed as evidence for the
neutralization of the net current. The $\delta$ for
other sunspots have also been computed and are given in Table 1.

The map of vertical current density j$_z$ for the same sunspot is
shown with intensity scale in the left panel of Figure 4.
The values are expressed in Giga Amperes per square meter (GA/m$^2$).
We can see that the distribution of j$_z$ is dominated by high amplitude
fluctuations on small scale as also reported in \cite{tiw09a}.
It is therefore difficult to make out any systematic behaviour of the
sign of j$_z$ as a function of $\varpi$.

The right panel of the same figure shows the total current within a
radius $\varpi$ as a function of $\varpi$.
As expected from the trend in Figure 3,
the total current shows evidence for a rapid decline after reaching
a maximum. Similar trends were seen in other sunspots.
We have also plotted in right panel of Figure 4, the net
current as calculated by the derivative method (viz. summation
of current densities calculated as the local curl of B).
We do see a trend of neutralization, although the effect is less
pronounced because of the larger noise present in the
derivative method. We can also infer from the right panel of Figure 4
that the increments of net vertical current flowing through annular
portions of the sunspot do show a reversal in sign.

Table 1 shows the summary of results for all the sunspots analyzed.
Along with the power law index $\delta$ of $B_{\psi}$ decrease,
we have also shown the average deviation of the azimuth from the
radial direction (``twist angle''), as well as the SSA.
The average deviation of the azimuth is well correlated with
the SSA for nearly circular sunspots, but is not correlated
with SSA for more irregularly shaped sunspots.
Thus, SSA is a more general measure of the global
twist of sunspots, irrespective of their shape.

\section{Discussion and Conclusions}

It is well known for astrophysical plasmas, that the plasma
distorts the magnetic field and the curl of this distorted field
produces a current by Ampere's law \citep{parker79}.
Parker's (1996) expectation of net zero current in a sunspot was
chiefly motivated by the concept of a fibril structure for the sunspot
field. However, he also did not rule out the possibility of vanishing
net current for a monolithic field where the
azimuthal component of the vector field in a cylindrical geometry declines
faster than 1/$\varpi$. While it is difficult to detect fibrils using the
Zeeman effect notwithstanding the superior resolution of SOT on {\it Hinode},
the stability and accuracy of the measurements have allowed us to detect
the faster than 1/$\varpi$ decline of the azimuthal component of the
magnetic field, which in turn can be construed as evidence for the
confinement of the sunspot field by the external plasma. The resulting
pattern of curl {\bf B} appears as a drop in net current at the
sunspot boundary.

If this lack of net current turns out to be a general feature
of sunspot magnetic fields in the photosphere, then measurement of
helicity from a global average of the force-free parameter becomes
suspect. On the other hand, sunspots are evidently twisted
at photospheric levels, as seen from the non-vanishing average twist
angle as well as the SSA (Table 1).
Although the existence of a global twist in the absence of a net
current is possible for a monolithic sunspot
field \citep{baty00,arch04,fan04,aula05},
a fibril model of the sunspot field can accommodate a global
twist even without a net current \citep{parker96}.

The spatial pattern of current density in a sunspot (e.g., left panel of Figure 4)
is really a manifestation of the deformation of the magnetic
field ($\nabla\times\bf B$) by the forces applied by the plasma.
The Lorentz force exerted by the field on the plasma produces
an equal and opposite force by the plasma, thereby confining the
field. Thus our analysis actually shows the pattern of the forces
exerted by the plasma on the field. The sharp decline of the
azimuthal field with radial distance thus shows the confinement
of the sunspot magnetic field by the radial gradient of the
plasma pressure.

Theoretical understanding of the penumbral fine structure has
improved considerably in recent times \citep{thom02,weis04}.
The onset of a convective instability for magnetic
field inclination exceeding a critical value was proposed
by \cite{tilde03} and \cite{hurl02}. A bifurcation in the onset
\citep{ruck95} could explain other features like hysteresis
in the appearance of penumbra as a function of sunspot size.
Numerical simulation of magneto-convection also steadily
improved \citep{hein07,remp09a}, culminating in very realistic
production of penumbral field structure \citep{remp09b}.
It is possible, owing to the random and stochastic nature of
convective structures, that no net twist in the simulated spot field
would be produced by convection for negligible Coriolis force.
If so, it would be very interesting to simulate magneto-convection
in a twisted sunspot field. In this case, would the resulting fine
structure mimic the observed ``curly interlocking combed''
structure of the penumbral magnetic field? If not, we
must look elsewhere for explaining the ``curly interlocking combed''
structure. A twisted fibril bundle would then be a solution.
Recent examples of filamentary penumbral structures based on such
cluster models \citep{sola93,spru06,scha06} have also been
proposed.

\cite{parker96} also mentions the possibility of net currents in
the corona, continuing down to the height where the first cleaving
takes place. It would therefore be imperative to look for net
currents at higher reaches of the solar atmosphere.
This is very important because several theories of
flares \citep{melr95} and CME triggers \citep{forb91,klie06}
rely heavily on the existence of net currents
in the corona above the sunspots.

Future large ground based telescopes equipped with adaptive optics
and multi spectral line capabilities would go a long way in
addressing these issues. In the meantime, direct measurement of
the global twist of sunspots using parameters like the SSA should
serve as proxies for estimating the net currents of
active regions in the corona. The SSA will also be a better
parameter to base a fresh look at the hemispheric rule in
photospheric chirality.

\acknowledgments {\bf Acknowledgements}\\
We thank Professor Eugene N. Parker for very instructive comments
on an early version of the manuscript. His comments on the
interpretation of currents in astrophysical plasmas have been
particularly useful. The remarks of an anonymous referee have
enhanced the clarity of the presentation and improved the
understanding of the results. The contributions of the late
Professor Metcalf to the inversion software package is also
acknowledged.
Hinode is a Japanese mission developed and launched by ISAS/JAXA,
collaborating with NAOJ as a domestic partner, NASA and STFC (UK)
as international partners. Scientific operation of the Hinode
mission is conducted by the Hinode science team organized at ISAS/JAXA.
This team mainly consists of scientists from institutes in the partner
countries. Support for the post-launch operation is provided by
JAXA and NAOJ (Japan), STFC (U.K.), NASA (U.S.A.), ESA, and NSC (Norway).

\begin{table}
%\begin{deluxetable}
\caption{List of the active regions studied. The power index
$\delta$: the slope of decrease of $B_{\psi}$ value, the twist angle,
the signed shear angle (SSA) and other details of the sunspots are
given:\label{tbl-1}}
%\tabletypesize{\tiny{}}
%\tablewidth{8pt}
\footnotesize{}
%\small{}
%\tiny{}
\centering
\begin{tabular}{c c c c c c c}
\hline     %%for horizontal space
%\startdata
AR No.      &  Date of        & Slope         & Shear Angle   & Twist Angle                   & Position    & Hemispheric \\
(NOAA)      &  Observation    & $\delta$      &(SSA: deg)     & $(tan^{-1}(B_{\psi}/B_r)$: deg) &           & Helicity Rule \\
\hline                                                                                                                                %twist_angle
10969	    & 29 Aug 2007  	  & $7.514$       & $-4.488$	  & $-4.009$                  & S05W33(t)	    & No  \\               %$8.150$
10966	    & 07 Aug 2007	  & $4.349$       & $-5.120$	  & $-7.028$                  & S06E20(t)	    & No  \\
10963($-$)  & 12 Jul 2007  	  & $4.366$       & $-5.123$      & $41.873^{\sharp}$         & S06E14(t)       & No  \\               %$41.873$
10963($+$)  & 12 Jul 2007  	  & $4.210$       & $-4.495$      & $-5.112$                  & S06E14(t)	    & No	 \\
10961       & 02 Jul 2007     & $4.976$       & $-4.973$	  & $29.451^{\sharp}$         & S10W16(t)	    & No	 \\               %$31.562$
10960	    & 07 Jun 2007  	  & $3.267$       & $3.182$	      & $-24.012^{\sharp}$        & S07W03	        & Yes \\               %-25
10953	    & 29 Apr 2007  	  & $8.249$       & $-3.382$	  & $7.200^{\sharp}$          & S10E22(t)	    & No	 \\               %7.200
10944       & 03 Mar 2007     & $2.407$       & $-4.635$      & $-5.130$                  & S05W30(t)	    & No	 \\
10940	    & 01 Feb 2007  	  & $2.281$       & $-4.726$      & $-7.950$                  & S04W05          & No	 \\
10933       & 05 Jan 2007     & $9.584$       & $-2.283$      & $-2.689$                  & S04W01	        & No	 \\
10926       & 03 Dec 2006     & $2.750$       & $-1.538$      & $6.001^{\sharp}$          & S09W32(t)       & No  \\               %6.001
10923	    & 10 Nov 2006  	  & $3.175$       & $0.785$	      & $-9.010^{\sharp}$         & S05W30(t)	    & Yes \\               %-9.010
%\enddata
\hline
%(t) : \it transformed \\
%$\sharp$ : {\it twist angle for irregular sunspots}% does not fit to a cylindrical assumption and therefore gives incorrect values.}
%(c) :  \it circular \\
%$^*$Table 1: online material
\end{tabular}
\tablecomments {1. (t) : \it transformed \\
$-$ 2. $\sharp$ : {\it twist angle for irregular sunspots does not fit to a cylindrical assumption and therefore gives incorrect values.}}
%\end{deluxetable}
\end{table}

\begin{figure}
%\plottwo{f1a.eps}{f1b.eps}
\epsscale{.80}
\plotone{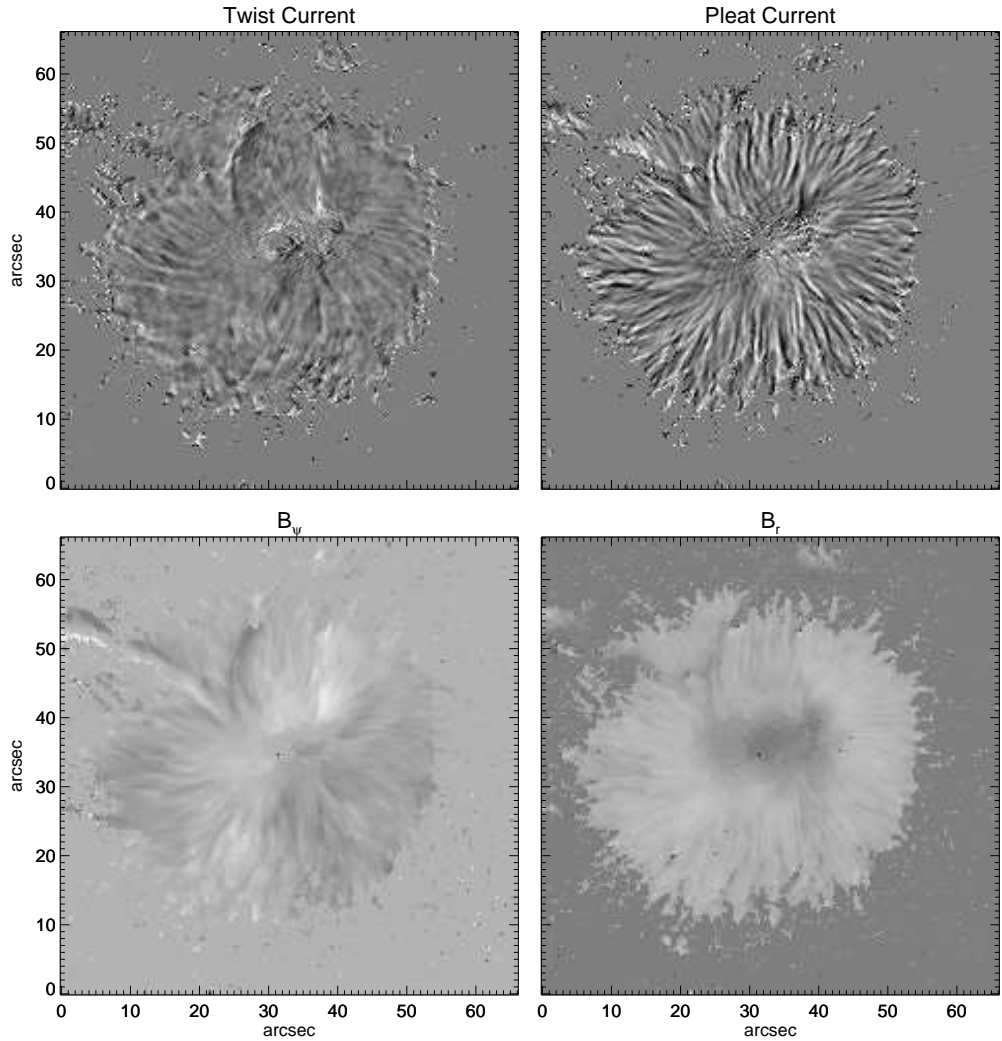}
\caption{Examples of the two components of the vertical electric current density
namely the ``twist'' and the ``pleat'' current densities ($j_t$ and $j_p$)
observed in NOAA AR 10933 are shown in the upper panel.
The lower panel shows the azimuthal
and radial component of the magnetic field ($B_{\psi}$ and $B_r$).}
\end{figure}

\begin{figure}
\epsscale{.80}
\plotone{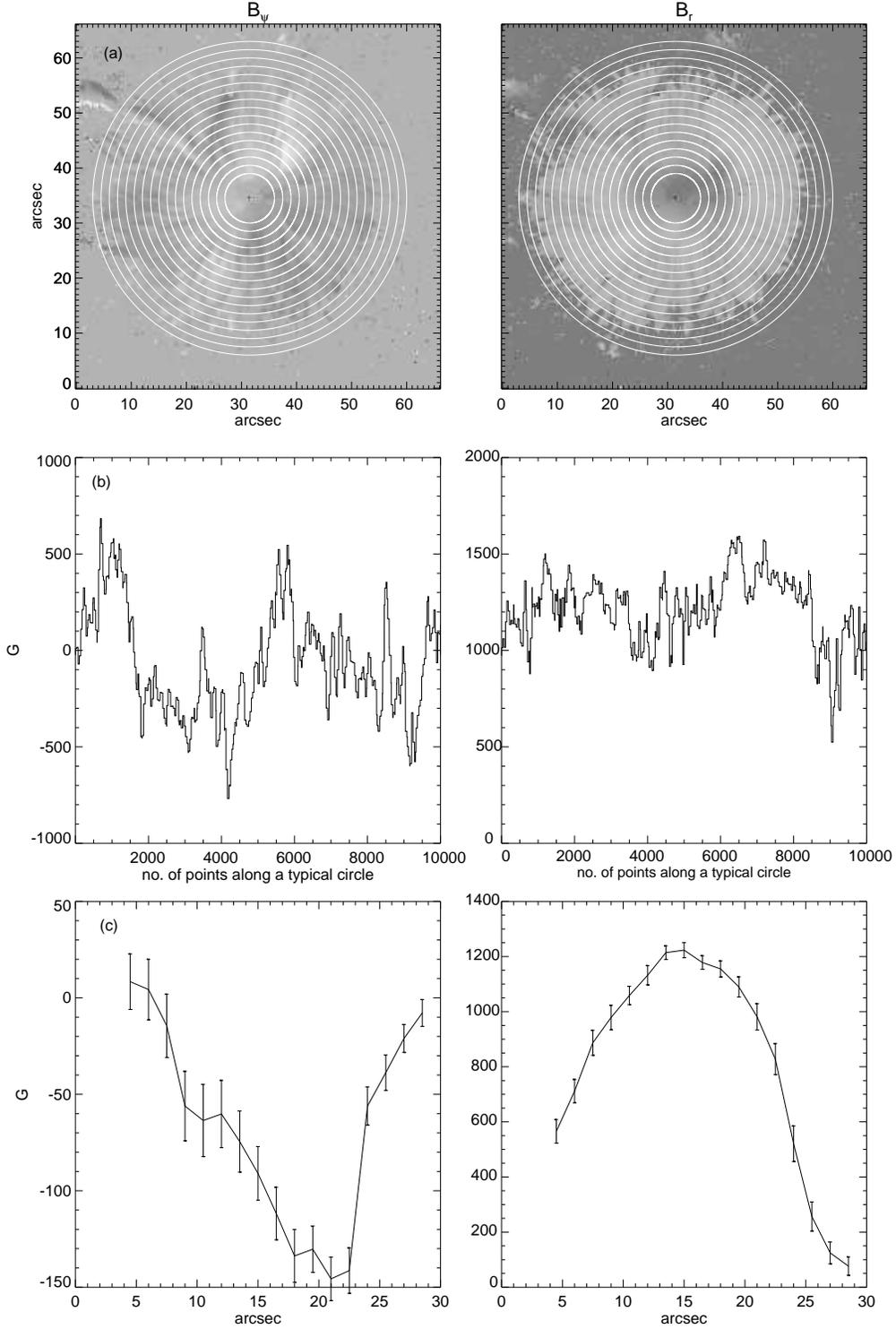}
\caption{(a): The same image as the lower panel of Figure 1 but with concentric circles
over plotted on them. (b): Plots of $B_{\psi}$ and $B_r$ along the periphery of
a typical circle (45$^{th}$ pixel away from center) selected in the sunspot.
(c): The mean $B_{\psi}$ and mean $B_r$ with $\frac{1}{16}$ and $\frac{1}{8}$
of their variations respectively along radial direction with different $\varpi$
have been plotted.}
\end{figure}

\begin{figure}
\epsscale{.50}
\plotone{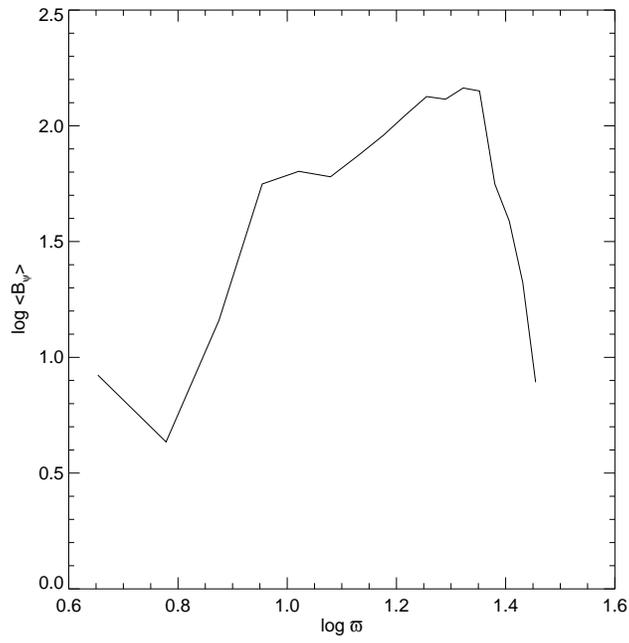}
\caption{Variation of log $<B_{\psi}>$ with log $\varpi$.
The slope of the declining portion of plot has been calculated as
a simple power law index $\delta$ for each sunspot and has been given
in Table 1.}
\end{figure}

\begin{figure}
\epsscale{1.0}
\plottwo{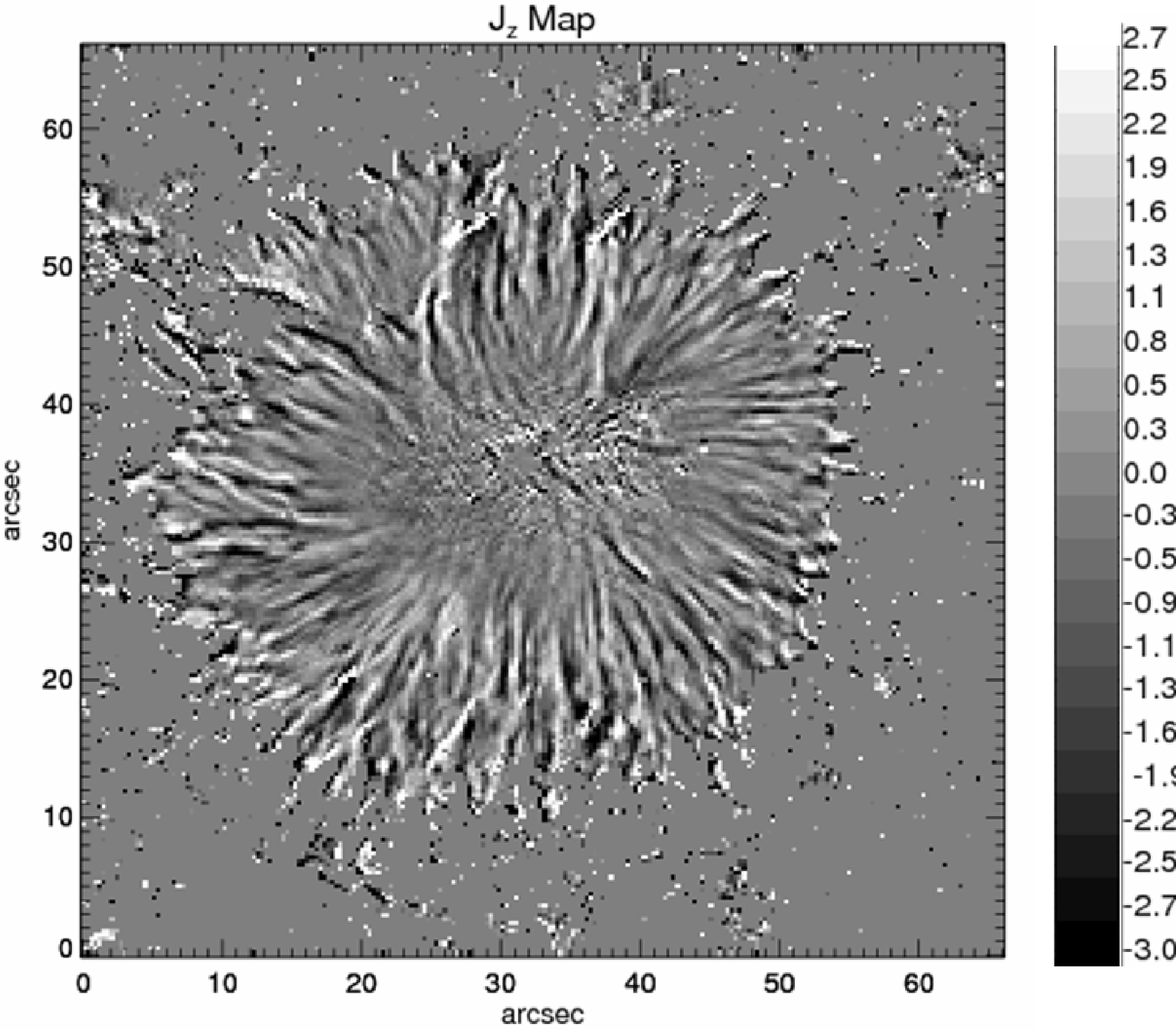}{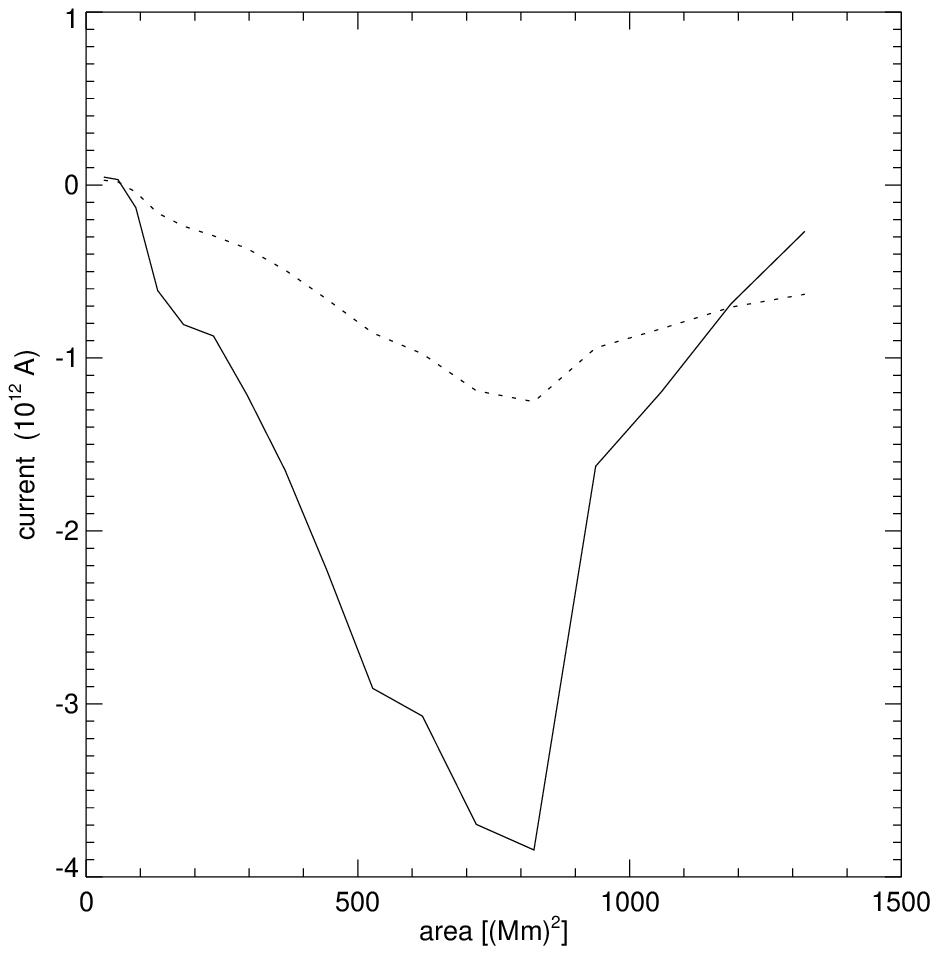}
\caption{Left panel: The map of vertical current density j$_z$ is shown with intensity
scale. The values are expressed in Giga Amperes per square meter (GA/m$^2$).
Right panel: The net current variation with increasing area has been shown. The solid
line shows the results of the calculations from the Equation 2. Also shown, by a
dashed line, is the results from the derivative method.
We can see the net current reduces very fast after a peak and almost vanishes
for complete sunspot. On the other hand the net current computed from the
derivative method shows a shallow behaviour.}
\end{figure}

%%%\twocolumn
%\bibliographystyle{apj}
%\bibliography{stiwari}

\end{document}